\documentclass[12pt,preprint]{aastex}

\shorttitle{Density Intermittency in ISM Turbulence}
\shortauthors{Terry \& Smith}

\begin{document}

\title{Coherence and Intermittency of Electron Density \\
  in Small-Scale Interstellar Turbulence}
  
\author{P.W. Terry and K.W. Smith}
\affil{Center for Magnetic Self Organization in Laboratory and Astrophysical Plasmas and Department of Physics, University of Wisconsin-Madison, Madison, WI 53706}
\email{pwterry@wisc.edu}

\begin{abstract}

Spatial intermittency in decaying kinetic Alfv\'{e}n wave turbulence is investigated to determine if it produces non Gaussian density fluctuations in the interstellar medium.  Non Gaussian density fluctuations have been inferred from pulsar scintillation scaling.  Kinetic Alfv\'{e}n wave turbulence characterizes density evolution in magnetic turbulence at scales near the ion gyroradius.  
It is shown that intense localized current filaments in the tail of an initial Gaussian probability distribution function possess a sheared magnetic field that strongly refracts the random kinetic Alfv\'{e}n waves responsible for turbulent decorrelation.  The refraction localizes turbulence to the filament periphery, hence it avoids mixing by the turbulence.  As the turbulence decays these long-lived filaments create a non Gaussian tail.  A condition related to the shear of the filament field determines which fluctuations become coherent and which decay as random fluctuations.  The refraction also creates coherent structures in electron density.  These structures are not localized. Their spatial envelope maps into a probability distribution that decays as density to the power $-3$.  The spatial envelope of density yields a Levy distribution in the density gradient.

\end{abstract}


\section{Introduction}

Pulasr radio signals probe fluctuations in the local interstellar medium \citep{arm81}.  The broad electron density fluctuation spectrum \citep{arm95} is commonly interpreted as a turbulent inertial range.  The pulsar signal width yields information about fluctuation statistics \citep{bhat04,sut71}.  The width scales as $R^4$ ($R$ is the distance to source), a result that  is incompatible with Gaussian statistics \citep{stas03a}.  The latter would produce a scaling of $R^2$, while $R^4$ is recovered for Levy statistics \citep{stas03b}.  A Levy-distributed random walk typically consists of a series of small random steps, punctuated by occasional Levy flights in which there is a single large jump to a new locale.  In the context of a pulsar radio signal propagating through a Levy distribution of electron density fluctuations, a sea of low intensity density fluctuations would scatter the signal through a series of small angles.  Intermittently, as the signal traversed an intense, localized density fluctuation, it would scatter through a much larger angle.

 The assertion that pulsar signals are dispersed by Levy-distributed fluctuations is a statistical ansatz validated to some degree by observation.  This ansatz does not address the difficult and important question of what processes or conditions produce the statistics.  It has been suggested that Levy statistics can emerge from radio signal trajectories grazing the surface of molecular clouds \citep{stas06}.  Here we examine a different mechanism rooted in the turbulent cascade implied by the broad fluctuation spectrum.  The mechanism is intrinsic spatial intermittency, a process known to create non Gaussian tails in the probability distribution function (PDF).  In Navier-Stokes turbulence, intrinsic intermittency takes the form of randomly dispersed, localized vortex strands, surrounded by regions of relative inactivity \citep{kerr85}.  Intermittency is most pronounced at small scales.  Intermittency also occurs in MHD turbulence \citep{grap91}.  However the statistical properties of electron density fluctuations in magnetic turbulence are not known.  In this paper we address the fundamental and nontrivial question of whether electron density can become intermittent in the magnetic turbulence of the interstellar medium.  The effect on pulse-width scaling requires that additional issues be addressed, and will be taken up later.  

The question of intermittency in pulsar scintillation is twofold.  First, can intermittent electron density fluctuations in interstellar turbulence achieve the requisite intensity to change the PDF?  To some extent this question has been answered by studies that show that passive advection and the limitations it places on electron density excitation (as indicated, for example, by mixing length arguments) applies only to scales larger than tens of gyroradii.  At smaller scales the electron density becomes active through kinetic Alfv\'{e}n wave (KAW) interactions with magnetic fluctuations, exciting the internal energy to equipartition with the magnetic energy \citep{ter01}.  Evidence for a transition to KAW dynamics near the gyroradius scale has recently been inferred from solar wind observations \citep{bale05}.  Since scintillation is dominated by small scales, the regime of kinetic Alfv\'{e}n interactions is appropriate for studying the intermittency potentially associated with the scaling of the pulsar signal width.  The second aspect of intermittency in the context of pulsar signals deals with how isolated structures can form against the homogenizing influence of turbulent mixing in a type of turbulence that does not involve flow.  Virtually all mechanisms proposed for intermittency involve flow or momentum, yet, the flow of ions in magnetic turbulence decouples from small-scale kinetic Alfv\'{e}n waves, with the interaction of magnetic field and density taking place against a background of unresponsive ions.

While intermittency has been widely studied in hydrodynamic turbulence \citep{fri95} and MHD \citep{pol95}, historically the emphasis has been on structure and statistics, not mechanisms.  Structure studies have included efforts to visualize intermittent structures \citep{head81}.  Quantitatively, measurements of structure function scalings have been made to gauge how intermittency changes with scale \citep{she94,pol95}.  Statistical characterizations of intermittency generally postulate a non Gaussian statistical ansatz, and the resultant properties are compared with measurements to determine if the ansatz is reasonable.  These approaches do not address the mechanisms that endow certain fluctuation structures either with individual longevity or collective prominence, in a statistical sense, relative to other regions in which such structures are not present \citep{wal97}.  The mechanistic approach is nascent but has already yielded significant insights into the long-standing problem of subcritical instability in plane Pouseille flow \citep{hof04}.

A starting point for our considerations are simulations of decaying KAW turbulence that showed the emergence of coherent, longlived current filaments under collisional dissipation of density \citep{crad91}.  
In these simulations finite amplitude fluctuations in density and magnetic field decayed from initial Gaussian distributions.  (The current, as curl of the magnetic field, was also Gaussian initially.)  The distribution of current became highly non Gaussian as certain current fluctuations persisted in the decay long past the nominal turbulent correlation time.  The longevity of these filaments enhanced the tail of the PDF, steadily increasing the value of the fourth order moment (kurtosis) significantly above its Gaussian value.  While the PDF was affected by mutual interactions of filaments later in the simulation, initially the tail enhancement was dominated by the interaction of filaments with surrounding turbulence, and the lack of mixing of those filaments relative to the rapidly-decaying surrounding turbulence.  Intermittency was not reported when resistivity dominated the dissipation.  
While these simulations showed intermittency in KAW turbulence, non Gaussian statistics was demonstrated for current fluctuations, not density.  The turbulence decayed via collisional dissipation of density; the current had no direct damping.  It is not clear what effect this had on density structure formation within the constraints of the resolution of the simulations.  The question of intermittency in density therefore remains open.  No mechanism for the intermittency was proposed.

In this paper we will examine analytically the dynamics of structures in density and current and determine how one relates to the other.  We will use analysis tools and results developed to understand the emergence of long lived vortices in decaying 2D Navier-Stokes turbulence \citep{mcw84}.  For that problem, two time-scale analysis showed that the vortices are coherent and long lived because strong shear flow in the outer part of the vortex suppresses ambient mixing by turbulence \citep{ter89,ter92}.  The ambient mixing would otherwise destroy the vortex in a turnover time.  This mechanism for maintaining the coherent vortex in decaying turbulence correctly predicts the observed distribution of Gaussian curvature of the flow field \citep{ter00}.

We use two-time-scale analysis to describe coherent structure formation in decaying KAW turbulence.  The following are obtained.  1) We identify the mechanism that allows certain current filaments to escape the turbulent mixing that otherwise typifies the turbulence.  Current and density are mixed by the random interaction of kinetic Alfv\'{e}n waves.  This process is disrupted in current filaments whose azimuthal field has an unusually large amount of transverse shear.  This creates a strong refraction of turbulent kinetic Alfv\'{e}n waves that localizes them to the periphery of the filament and restricts their ability to mix current and density.  2) We derive a shear threshold criterion based on this mechanism.  It identifies which current filaments escape mixing and become coherent, or long lived.  The criterion relates to the Gaussian curvature of the magnetic field, providing a topological construct that maps the intermittency in a way analogous to the flow Gaussian curvature of decaying 2D Navier-Stokes turbulence.  3) We trace the relative effects of the refractive shear mechanism on current, magnetic field, density, and flux.  The magnetic field and density have long-lived, localized fluctuation structures that coexist spatially with localized current filaments.  However, the magnetic field extends beyond the localized current.   Like the magnetic field of a line current, it falls off as $r^{-1}$.  Because the density is equipartitioned with the magnetic field in KAW turbulence a similar mantle is expected for the density.  This mantle tends to prevent the density kurtosis from rising to values greatly above 3; however, it is responsible for giving the PDF of density gradient a Levy distribution.

This paper is organized as follows.  Section II presents the kinetic Alfv\'{e}n wave model used in this paper.  The two-time-scale analysis is introduced in Sec.~III.  Section IV derives the condition for strong refraction, and the resultant refractive boundary-layer structure for turbulent KAW activity in and around the coherent filament.  The turbulent mixing stresses are determined in Sec.~V, from which the filament and density lifetimes can be derived.  Section VI discusses the Gaussian curvature and spatial properties of the current and density structures.  The latter are used to infer heuristic PDFs.  Conclusions are given in Sec.~VII.

\section{Kinetic Alfv\'{e}n Wave Model}

The shear-Alfv\'{e}n and kinetic-Alfv\'{e}n physics described in the introduction is intrinsic to models of MHD augmented by electron continuity.  When there is a strong mean field, the nonlinear MHD dynamics can be represented by a reduced description \citep{haz83}, given by  

\begin{equation}
\frac{\partial \hat{\psi}}{\partial t}+\nabla_{||}\hat{\phi}=\eta \hat{J}+\nabla_{||}\hat{n}+\frac{C_s}{V_A}\frac{1}{n_0}{\bf \nabla}\hat{\psi}\times{\bf z}\cdot{\bf \nabla}n_0,
\end{equation}
\begin{equation}
\frac{\partial}{\partial t} \nabla_{\perp}^2\hat{\phi}-{\bf \nabla}\hat{\phi}\times{\bf z} \cdot{\bf \nabla}\nabla_{\perp}^2\hat{\phi}=-\nabla_{||}\hat{J},
\end{equation}
\begin{equation}
\frac{\partial \hat{n}}{\partial t}-{\bf \nabla}\hat{\phi}\times{\bf z} \cdot{\bf \nabla}\hat{n}+\nabla_{||}\hat{J}-\frac{C_s}{V_A}\frac{1}{n_0}{\bf \nabla}\hat{\phi}\times{\bf z}\cdot{\bf \nabla}n_0=0,
\end{equation}
\noindent{where}
\begin{equation}
\nabla_{||}=\frac{\partial}{\partial z}+{\bf \nabla}\hat{\psi}\times{\bf z}\cdot{\bf \nabla},
\end{equation}

\noindent{and} $\hat{J}=\nabla_{\perp}^2\hat{\psi}=\partial^2\hat{\psi}/\partial x^2+\partial^2\hat{\psi}/\partial y^2$.  In the reduced description the perturbed magnetic field is perpendicular to the mean field and can be written as $\hat{b}/B={\bf \nabla}\hat{\psi}\times{\bf z}$, where $z$ is the direction of the mean field, and $\hat{\psi}=(C_s/c)eA_z/T_e$ is the normalized parallel component of the vector potential.  The flow has zero mean and is also perpendicular to the mean field $B$.   It can be expressed in terms of the electrostatic potential as  $-{\bf \nabla}\hat{\phi}\times{\bf z}$ where $\hat{\phi}=(C_s/V_A)e\phi/T_e$ is the normalized electrostatic potential.  The normalized density fluctuation is $\hat{n}=(C_s/V_A)\tilde{n}/n_0$, where $n_0$ is the mean density, and $\eta=(c^2/4\pi V_A\rho_s)\eta_{sp}$ is the normalized resistivity, where $\eta_{sp}$ is the Spitzer resistivity.  Spatial scales are normalized to $\rho_s=C_s/\Omega_i$, time is normalized to the Alfv\'{e}n time $\tau_A=\rho_s/V_A$, $C_s=(T_e/m_i)^{1/2}$ is the ion acoustic velocity, $V_A=B/(4\pi m_in_0)^{1/2}$ is the Alfv\'{e}n velocity, and $\Omega_i=eB/m_ic$ is the ion gyrofrequency.  Within their limitations (isothermal, incompressible fluctuations), Eqs.~(1)-(4) are valid for scales both large and small compared to the gyroradius, as well as the intermediate region.

Equation (3) is the electron continuity equation.  The advective nonlinearity, ${\bf \nabla}\hat{\phi}\times{\bf z}\cdot{\bf \nabla}\hat{n}$, couples electron density fluctuations to the flow.  If there is a nonuniform mean density, advection drives weak density fluctuations of amplitude $\hat{n}\approx(\delta/L_n)n_0$, where $\delta$ is the scale of density fluctuations and $L_n$ is the mean density scale length.   The continuity equation also contains a compressible nonlinearity, ${\bf \nabla}_{||}\hat{J}$, whereby compressible electron motion along magnetic field perturbations provides coupling to the magnetic field.  Electrons act on the magnetic field through parallel electron pressure in Ohm's law, expressed as ${\bf \nabla}_{||}\hat{n}$ in Eq.~(1).  The couplings of magnetic field and density are weak at scales appreciably larger than the ion gyroradius.  On those scales the advection of electron density is passive to a good approximation, and governs electron density evolution.  In the region around $\delta=10\rho_s$, the two nonlinearities in each of Eqs.~(1)-(3) become comparable \citep{ter01}.
For $\delta<10\rho_s$, ${\bf \nabla}_{||}\hat{n}$ begins to dominate ${\bf \nabla}_{||}\hat{\phi}$ in Eq.~(1), and ${\bf \nabla}_{||}\hat{J}$ begins to dominate ${\bf \nabla}\hat{\phi}\times{\bf z}\cdot{\bf \nabla}\hat{n}$ in Eq.~(3).  This is a very different regime from incompressible MHD, where the magnetic field and flow actively exchange energy through shear Alfv\'{e}n waves.  In a turbulent cascade approaching the ion gyroradius scale from larger scales, the energy exchanged between flow and magnetic field in shear Alfv\'{e}n interactions diminishes relative to the energy exchanged between the electron density and the magnetic field through the compressible coupling.  Consequently flow decouples from the magnetic field, increasingly evolving as a go-it-alone Kolmogorov cascade, while electron density and magnetic field, interacting compressively through kinetic Alfv\'{e}n waves, supplant the shear Alfv\'{e}n waves.   Once the kinetic Alfv\'{e}n wave coupling reaches prominence, the internal and magnetic energies become equipartitioned, $\int \hat{n}^2dV\approx \int |{\bf \nabla}\hat{\psi}|^2dV$, even if the internal energy  is only a fraction of the magnetic energy at larger scales.  If there is no significant damping at the ion gyroradius scale, the large-scale shear Alfv\'{e}n cascade continues to gyroradius scales and beyond though kinetic Alfv\'{e}n waves.  The gyroradius scale at which KAW dynamics is active is order $10^8$ cm in the warm ISM.  This is small relative to the scale of intermittent flow structures in molecular gas clouds, recently reported to be order $10^{18}$ cm \citep{hily07}.  This scale difference is crudely consistent with the high magnetic Prandtl number of the warm ISM.  The value Pr $\sim10^{14}$ allows very small scales in the ionized medium, before dissipation becomes important, relative to scales of viscous dissipation in the clouds.
The gyroradius scale of intermittent KAW structures makes direct visualization in the ISM difficult. 

In the KAW regime, the model can be further simplified by dropping the flow evolution.  This leaves a KAW model in which electron density and magnetic field interact against a neutralizing background of unresponsive ions, 

\begin{equation}
\frac{\partial \hat{\psi}}{\partial t}=\frac{\partial \hat{n}}{\partial z}+{\bf \nabla}\hat{\psi}\times{\bf z}\cdot{\bf \nabla}\hat{n}+\eta \hat{J}+\frac{C_s}{V_A}\frac{1}{n_0}{\bf \nabla}\hat{\psi}\times{\bf z}\cdot{\bf \nabla}n_0,
\end{equation}
\begin{equation}
\frac{\partial \hat{n}}{\partial t}+\frac{\partial \hat{J}}{\partial z}+{\bf \nabla}\hat{\psi}\times{\bf z}\cdot{\bf \nabla}\hat{J}-\mu\nabla^2\hat{n}=0.
\end{equation}

\noindent{Solutions} of this model closely approximate those of Eqs.~(1)-(3) when the scales are near the gyroradius or smaller \citep{ter98}.  This model assumes isothermal fluctuations, consistent with strong parallel thermal conductivity.  Equations (5) and (6) are fluid equations, hence Landau-resonant \citep{how06} and gyro-resonant dissipation, which may be important in the ISM, are not modeled.  Ohm's law has resistive dissipation, and density evolution has collisional diffusion.  Depending on the ratio $\eta/\mu$, either of these dissipation mechanisms can damp the energy in decaying turbulence, however, the damping occurs at small dissipative scales.  We will focus on inertial behavior at larger scales.  We assume that mean density is nearly uniform, and neglect the last term of Eq.~(5).

The dispersion relation for ideal kinetic Alfv\'{e}n waves is determined by linearizing Eqs.~(5) and (6), neglecting resistive dissipation $\eta\hat{J}$, and introducing a Fourier transform in space and time.  The result is $\omega=k_zk_{\perp}$,
where $\textbf{k}_{\perp}\cdot\textbf{z}=0$, or $\textbf{k}_{\perp} \perp k_z\textbf{z}$. If dimensional frequency and wavenumbers $\tilde{\omega}$, $\tilde{k}_z$, and $\tilde{k}_{\perp}$ are reintroduced, the dispersion relation is $\tilde{\omega}=V_A\tilde{k}_z\tilde{k}_{\perp}\rho_s$.  The wave is seen to combine Alfv\'{e}nic propagation with perpendicular motion associated with the gyroradius scale.  The KAW eigenvector yields equal amplitudes of magnetic field and the density, $-k_{\perp}\psi_k=ib_k=n_k$,
with a phase difference of $\pi/2$.

In magnetic turbulence with its hierarchy of scales, kinetic Alfv\'{e}n waves also propagate along components of the turbulent magnetic field.  In the reduced description the turbulent field is perpendicular to the mean field, hence the dispersion relation of these kinetic Alfv\'{e}n waves carries no $k_z$ dependence.  To illustrate, we isolate such a fluctuation from the mean-field kinetic Alfv\'{e}n wave by setting $k_z=0$; with this wavenumber zero, we drop the subscript from $k_{\perp}$; we consider a turbulent magnetic field component $\hat{\textbf{b}}_{k_0}/B_0=i\textbf{k}_0\times\textbf{z}\psi_{k_0}$ at wavenumber $k_0$ that dominates the low-$k$ fluctuation spectrum; and we look at the dispersion for smaller scale fluctuations satisfying $k\gg k_0$.  The latter conditions linearize the problem, yielding a dispersion relation for kinetic Alfv\'{e}n waves propagating along the turbulent field $\hat{\textbf{b}}_{k_0}$ according to $\omega=i\textbf{k}_0\times\textbf{z}\cdot\textbf{k}\textrm{ }\psi_{k_0}k=[\hat{\textbf{b}}_{k_0}\cdot\textbf{k}/B]\textrm{ }k$.
Reintroducing dimensions, $\tilde{\omega}=V_A(\textbf{b}_{k_0}\cdot\tilde{\textbf{k}}/B)\tilde{k}\rho_s$.   We see that the dispersion is Alfv\'{e}nic, but with respect to a perturbed field component that is perpendicular to the mean field.  Hence the frequency goes like $\tilde{k}^2\rho_s$ instead of $\tilde{k}_z\tilde{k}\rho_s$.

\section{Two Time Scale Analysis}

To understand and quantify the conditions under which a coherent current fluctuation persists for long times relative to typical fluctuations, we examine the interaction of the coherent structure with surrounding turbulence and derive its lifetime under turbulent mixing.  The interaction is described using a two-time-scale analysis, allowing evolution on disparate time scales to be tracked \citep{ter92}.  The coherent structure, a current filament with accompanying magnetic field and electron density fluctuations, evolves on the slow time scale under the rapid-scale-averaged effect of turbulence.  On the rapid scale the filament is essentially stationary, creating an inhomogeneous background for the rapidly evolving turbulence.  Identifying conditions that support longevity justifies the two time scale approximation a posteriori.  
Simulations suggest the filament is roughly circular.  If coordinates are chosen with the origin at the center of the filament, a circular filament is azimuthally symmetric, while the turbulence breaks that symmetry.

The filament current is localized, hence its current density becomes zero at some distance from the origin.  The localized current profile necessarily creates a magnetic field that is strongly inhomogeneous.  On the rapid time scale over which the turbulence evolves, this field, which is part of the coherent structure, is essentially stationary.  It acts as a secondary equilibrium field in addition to the primary equilibrium field (which is homogeneous and directed along the $z$-axis).  Turbulence, in the form of random kinetic Alfv\'{e}n waves, propagates along both the primary and secondary fields.  Because the primary field is homogeneous its effect on the turbulence is uninteresting.  However, the secondary field is strongly sheared because of the local inhomogeneity created by the structure.  Strong shear refracts the turbulent kinetic Alfv\'{e}n waves.  In the subsequent analysis we will ignore the primary KAW propagation, which we can do by setting $k_z=0$, and focus on the refraction of KAW propagation by the secondary magnetic field shear.  Strong refraction will be shown to localize kinetic Alfv\'{e}n waves away from the heart of the filament, allowing it to escape mixing and thereby acquire the longevity to make it coherent.

With $J=\nabla_{\perp}^2\hat{\psi}$, we apply the separation of long and short time scales to fluctuations in $\hat{n}$ and $\hat{\psi}$ as follows:
\begin{equation}
\hat{F}=F_0(r,\tau)+\tilde{F}(r,\theta,t),
\end{equation}
\noindent{where} the symbol $F$ represents either $\psi$ or $n$, with $\psi_0$ and $n_0$ the flux function and density of the coherent structure, and $\tilde{\psi}$ and $\tilde{n}$ the turbulent fields of flux and density.  The variables for slowly and rapidly evolving time are $\tau$ and $t$.  The origin of a polar coordinate system with radial and angle variables $r$ and $\theta$ is placed at the center of the structure.  The structure is assumed to be azimuthally symmetric. The turbulence evolves in the presence of the structure, hence it is necessary to specify the radial profile of $\psi_0$, or more explicitly, the profile of the secondary, structure field $\nabla\psi_0(r)\times\hat{{\bf z}}=B_{\theta}(r)\hat{\theta}$.  As a generic profile for localized current we adopt a reference profile that peaks at the origin and falls monotonically to zero over a finite radius $a$.  For simplicity we take the variation as quadratic, giving
\begin{equation}
J_0(r)=J_0(0)\Big(1-\frac{r^2}{a^2}\Big),
\end{equation}

\noindent{where} $J_0(r)=\partial^2\psi_0/\partial r^2$.  Integrating the current we obtain
\begin{eqnarray}
B_{\theta}(r)=\frac{J_0(0)}{2}r\Big(1-\frac{r^2}{2a^2}\Big)& &(r\leq a), \label{interiorB} \\ [3mm]
B_{\theta}(r)=\frac{J_0(0)a^2}{4r}& &(r\geq a). 
\end{eqnarray}
These profiles are for reference.  Shortly we will introduce a more general description for a filament whose current peaks at the origin and decays monotonically.  The current of the coherent filament is wholly localized within $r=a$.  However, the magnetic field is not localized, but slowly decays as $r^{-1}$ outside the filament.  The quantities in Eqs.~(8)-(10) all evolve on the slow time scale $\tau$.  The dependence on $\tau$ is not notated because when $B_{\theta}$ appears in the turbulence equations, it is a quasi equilibrium quantity on the rapid time scale.

To describe the rapid time scale evolution and its azimuthal variations we introduce a Fourier-Laplace transform,
\begin{equation}
\tilde{F}(r,\theta,t)=\frac{1}{2\pi i}\int_{-i\infty+\gamma_0}^{+i\infty+\gamma_0}d\gamma\sum_m\tilde{F}_{m,\gamma}(r)\exp[-im\theta]\exp[\gamma t],
\end{equation}

\noindent{where} $\gamma_0$ is the shift of the complex integration path of the inverse Laplace transform.  The radial variation of $B_{\theta}(r)$ creates an inhomogeneous background field for the turbulence, making Fourier transformation unsuitable for the radial variable.  The Laplace transform is appropriate for turbulence that decays from an initial state.  To obtain equations for the slowly evolving fields $\psi_0$ and $n_0$, we average the full equations over the rapid time scale $t$.  This is accomplished by applying the Laplace transform to the equations, and integrating over $t$.  The integral selects $\gamma=0$ as the time average, {\it i.e.}., $\int dt f(\tau,t)=\int dt \int d\gamma(2\pi i)^{-1}f_{\gamma}(\tau)\exp(\gamma t) =f_{\gamma=0}(\tau)$.  Applying this procedure, the evolution equations for the slowly evolving fields are given by

\begin{equation}
\frac{\partial\psi_{\gamma=0}(\tau)}{\partial \tau}-\frac{1}{2\pi i}\int_{-i\infty+\gamma_0}^{-i\infty+\gamma_0}d\gamma'\sum_{m'}\Bigg\langle\Bigg[\tilde{b}_{r(-m',-\gamma')}\frac{\partial}{\partial r}+\tilde{b}_{\theta(-m',-\gamma')}\Big(\frac{im'}{r}\Big)\Bigg]\tilde{n}_{m',\gamma'}\Bigg\rangle=0,
\end{equation}

\begin{eqnarray}
\frac{\partial n_{\gamma=0}(\tau)}{\partial \tau}&+&\frac{1}{2\pi i}\int_{-i\infty+\gamma_0}^{-i\infty+\gamma_0}d\gamma'\sum_{m'}\Bigg\langle\Bigg[\tilde{b}_{r(-m',-\gamma')}\frac{\partial}{\partial r}+\tilde{b}_{\theta(-m',-\gamma')}\Big(\frac{im'}{r}\Big)\Bigg]\times \nonumber \\ [3mm]  
&&\Bigg[\frac{1}{r}\frac{\partial}{\partial r}\Big(r\frac{\partial}{\partial r}\Big)-\frac{m'^2}{r^2}\Bigg]\tilde{\psi}_{m',\gamma'}\Bigg\rangle=0, 
\end{eqnarray}
where $\tilde{b}_{r(-m',-\gamma')}$, $\tilde{b}_{\theta(-m',-\gamma')}$, $n_{m',\gamma'}$, and $\tilde{\psi}_{m',\gamma'}$ are understood to depend on the radial variable $r$; $\tilde{b}_{r(-m',-\gamma')}=(-im'/r)\psi_{-m',-\gamma'}$; and $\tilde{b}_{\theta(-m',-\gamma')}=-(\partial/\partial r)\psi_{-m',-\gamma'}$.  The correlations $\langle \tilde{b}_r\partial\tilde{n}/\partial r\rangle$, $\langle \tilde{b}_{\theta}\tilde{n}\rangle$,  $\langle \tilde{b}_r\partial\nabla^2\tilde{\psi}/\partial r\rangle$,  and $\langle \tilde{b}_{\theta}\nabla^2\tilde{\psi}\rangle$, which appear in Eqs.~(12) and (13), are turbulent stresses associated with random kinetic Alfv\'{e}n wave refraction.  Their fast time averages govern the mixing (transport) of the coherent fields.  These stresses must be evaluated from solutions of the fast time scale equations to find the lifetime of the structure.

The evolution equations for the rapidly evolving turbulent fluctuations are
\begin{eqnarray}
\gamma\tilde{\psi}_{m,\gamma}-B_{\theta}(r)\Big(\frac{-im}{r}\Big)\tilde{n}_{m,\gamma}+\frac{1}{2\pi i}\int_{-i\infty+\gamma_0}^{i\infty+\gamma_0}d\gamma'\sum_{m'}&\times& \nonumber \\ [2mm]
\Bigg[\frac{im'}{r}\tilde{\psi}_{m',\gamma'}\frac{\partial}{\partial r}-\frac{i(m-m')}{r}\frac{\partial \tilde{\psi}_{m',\gamma'}}{\partial r}\Bigg]\tilde{n}_{m-m',\gamma-\gamma'}&=&\frac{im}{r}\tilde{\psi}_{m,\gamma}\frac{\partial}{\partial r}n_0(r),
\end{eqnarray}

\begin{eqnarray}
&&\gamma\tilde{n}_{m.\gamma}-B_{\theta}(r)\Big(\frac{-im}{r}\Big)\Bigg[\frac{1}{r}\frac{\partial}{\partial r}\Big(r\frac{\partial}{\partial r}\Big)-\frac{m^2}{r^2}\Bigg]\tilde{\psi}_{m,\gamma}-\frac{1}{2\pi i}\int_{-i\infty+\gamma_0}^{i\infty+\gamma_0}d\gamma'\sum_{m'}\times \nonumber \\ [2mm]
&&\Bigg[\frac{im'}{r}\tilde{\psi}_{m',\gamma'}\frac{\partial}{\partial r}-\frac{i(m-m')}{r}\frac{\partial \tilde{\psi}_{m',\gamma'}}{\partial r}\Bigg]\Bigg[\frac{1}{r}\frac{\partial}{\partial r}\Big(r\frac{\partial}{\partial r}\Big)-\frac{(m-m')^2}{r^2}\Bigg]\tilde{\psi}_{m-m',\gamma-\gamma'} \nonumber \\ [2mm]
&=&\frac{im}{r}\tilde{\psi}_{m,\gamma}\frac{\partial}{\partial r} J_0(r).
\end{eqnarray} 

\noindent{We} have not shown the dissipative terms in accordance with our focus on inertial scales.  The sources contain gradients of $n_0(r)$ and $J_0(r)$.  These are the density and current of the coherent structure, but, unlike $\psi_{\gamma=0}(\tau)$ and $n_{\gamma=0}(\tau)$, are not evaluated in the Laplace transform domain.  Three terms drive the evolution of $\partial\tilde{\psi}/\partial t$ and  $\partial\tilde{n}/\partial t$ in each of these equations.  The first term describes linear kinetic Alfv\'{e}n wave propagation along the inhomogeneous secondary magnetic field $B_{\theta}$ of the coherent structure.  The second term is the nonlinearity, and describes turbulence of random kinetic Alfv\'{e}n waves.  The third term is proportional to mean-field gradients.  It is a fluctuation source via the magnetic analog of advection  ($\nabla\hat{\phi}\times\hat{{\bf z}}\cdot\nabla={\bf v}\cdot\nabla$  $\rightarrow$  $\nabla\hat{\psi}\times\hat{{\bf z}}\cdot\nabla$).  It yields quasilinear diffusivities for the turbulent mixing process.  For example, if the kinetic Alfv\'{e}n wave and nonlinear terms of Eq.~(15) are dropped, the solution is
\begin{equation}
 \tilde{n}_{m,\gamma}^{(i)}=\frac{1}{\gamma}\frac{im}{r}\tilde{\psi}_{m,\gamma}\frac{\partial J_0(r)}{\partial r}.
 \label{linearn}
 \end{equation} 
The superscript $(i)$ indicates that, for deriving diffusivities, this density is to be substituted iteratively into the correlations of the turbulent stresses.   From Eq.~(12) these correlations are $\langle \tilde{b}_r\tilde{n}\rangle=\langle \tilde{b}_r^{(i)}\tilde{n}\rangle+\langle \tilde{b}_r\tilde{n}^{(i)}\rangle$ and $\langle \tilde{b}_{\theta}\partial\tilde{n}/\partial r\rangle=\langle \tilde{b}_{\theta}^{(i)}\partial\tilde{n}/\partial r\rangle+\langle \tilde{b}_{\theta}\partial\tilde{n}^{(i)}/\partial r\rangle$.  Substitution of Eq.~(16) yields mean turbulent diffusivities for $\psi_0$.  Similarly, if Eq.~(14) is solved by dropping its kinetic Alfv\'{e}n wave and nonlinear terms, we obtain
 \begin{equation}
  \tilde{\psi}_{m,\gamma}^{(i)}=\frac{1}{\gamma}\frac{im}{r}\tilde{\psi}_{m,\gamma}\frac{\partial n_0(r)}{\partial r}.
  \label{linearpsi}
 \end{equation} 
Substituting this solution into the correlations $\langle \tilde{b}_r\nabla^2\tilde{\psi}\rangle=\langle \tilde{b}_r^{(i)}\nabla^2\tilde{\psi}\rangle+\langle \tilde{b}_r\nabla^2\tilde{\psi^{(i)}}\rangle$ and $\langle \tilde{b}_{\theta}\partial\nabla^2\tilde{\psi}/\partial r\rangle=\langle \tilde{b}_{\theta}^{(i)}\partial\nabla^2\tilde{\psi}/\partial r\rangle+\langle \tilde{b}_{\theta}\partial\nabla^2\tilde{\psi}^{(i)}/\partial r\rangle$ of Eq.~(13), mean turbulent diffusivities are obtained for $n_0$.  Off-diagonal transport (relaxation of $\psi_0$ by gradient of $n_0$) can also be obtained by substituting Eq.~(17) into $\langle \tilde{b}_r\tilde{n}\rangle$ and $\langle \tilde{b}_{\theta}\partial\tilde{n}/\partial r\rangle$.  The role of the nonlinear and kinetic Alfv\'{e}n wave terms omitted from Eqs.~(16) and (17) is to modify the time scale $\gamma$ and couple the sources.  This is calculated in the next section.  The inverse of $\gamma$ represents the lifetime of the correlations $\langle \tilde{b}_r\tilde{n}\rangle$, $\langle \tilde{b}_{\theta}\partial\tilde{n}/\partial r\rangle$,  $\langle \tilde{b}_r\nabla^2\tilde{\psi}\rangle$,  and $\langle \tilde{b}_{\theta}\partial\nabla^2\tilde{\psi}/\partial r\rangle$.  Generally the nonlinear terms enhance decorrelation, increasing the effective value of $\gamma$.  If the shear in $B_{\theta}$ is strong, the kinetic Alfv\'{e}n wave term increases $\gamma$ even further.

The role of shear in the kinetic Alfv\'{e}n wave terms is not explicit but should be, so that it can be varied independently of the field amplitude $B_{\theta}(r_0)$ at some radial location $r_0$.  In explicitly displaying the role of shear we note that if $B_{\theta}(r)\sim r$, as would be true if the current density $J_0$ were uniform, the kinetic Aflv\'{e}n wave term is independent of $r$.  In this situation the phase fronts of kinetic Alfv\'{e}n waves propagating along $B_{\theta}$ are straight-line rays extending from the origin.  Shear in $B_{\theta}$, occurring through nonuniformity of $J_0$, distorts the phase fronts, as shown in Fig. 1.  Distortion occurs if $B_{\theta}$ has a variation that is not linear.  From Eq.~(9) we note that the variation of $B_{\theta}$ for our chosen structure profile is linear for $r\ll a$, with variations developing as $r\rightarrow a$.  Therefore, it makes sense to quantify the shear by expanding $B_{\theta}(r)/r$ in a Taylor series about some point of interest.  Obviously, the shear is zero at the origin, and becomes sizable as $r\rightarrow a$.  Expanding about a reference point $r_0$ away from the origin, 
\begin{equation}
\frac{B_{\theta}(r)}{r}=\frac{B_{\theta}(r_0)}{r_0}+(r-r_0)\frac{d}{dr}\Big(\frac{B_{\theta}}{r}\Big)\Big|_{r_0}+\textrm{ . . . .  }.
\end{equation}
If $B_{\theta}(r)$ varies smoothly, as is the case for a monotonically decreasing current profile, we can truncate the expansion as indicated in Eq.~(18) and use that expression as a general current profile.  Looking at the kinetic Alfv\'{e}n terms of Eqs.~(14) and (15), the first term will produce a uniform frequency that Doppler shifts $\gamma$ by the amount $imB_{\theta}(r_0)/r_0$.  The second term will describe KAW propagation in an inhomogeneous medium with its attendant refraction.

\section{Refraction Boundary Layer}

We rewrite Eq.~(15), substituting the expansion of Eq.~(18), yielding
\begin{eqnarray}
\hat{\gamma}\tilde{n}_{m,\gamma}-im(r-r_0)\frac{d}{dr}\Big(\frac{B_{\theta}}{r}\Big)\Big|_{r_0}\nabla_m^2\tilde{\psi}_{m,\gamma}+\frac{1}{2\pi i}\int_{-i\infty+\gamma_0}^{i\infty+\gamma_0}d\gamma'\sum_{m'}&\times& \nonumber \\ [2mm]
\Bigg[\frac{im'}{r}\tilde{\psi}_{m',\gamma'}\frac{\partial}{\partial r}-\frac{i(m-m')}{r}\frac{\partial \tilde{\psi}_{m',\gamma'}}{\partial r}\Bigg]\nabla_{m-m'}^2\tilde{\psi}_{m-m',\gamma-\gamma'}&=&\frac{im}{r}\tilde{\psi}_{m,\gamma}\frac{\partial}{\partial r}J_0(r),
\end{eqnarray}
where
\begin{equation}
\nabla_m^2=\frac{1}{r}\frac{\partial}{\partial r}\Big(r\frac{\partial}{\partial r}\Big)-\frac{m^2}{r^2}
\end{equation}
and $\hat{\gamma}=\gamma+imB_{\theta}(r_0)/r_0$.  When $d/dr[B_{\theta}/r]|_{r_0}$ is large, the shear in $B_{\theta}$ refracts turbulent KAW activity.  The process can be described using asymptotic analysis.  In the limit that $d/dr[B_{\theta}/r]|_{r_0}$ becomes large asymptotically, the higher derivative nonlinear term is unable to remain in the dominant asymptotic balance unless the solution develops a small scale boundary layer structure.  The layer is a singular structure.  Its width must become smaller as $d/dr[B_{\theta}/r]|_{r_0}$ becomes larger, otherwise the highest order derivative drops out of the balance and the equation changes order.  This is the only viable asymptotic balance for $d/dr[B_{\theta}/r]|_{r_0}\rightarrow\infty$.  The boundary layer width $\Delta r$ is readily estimated from dimensional analysis by noting that $r-r_0\sim\Delta r$, $\partial\tilde{n}_m(t)/\partial r\sim\tilde{n}_m/\Delta r$, and treating $d/dr[B_{\theta}/r]|_{r_0}\equiv j'$ as the diverging asymptotic parameter.  The asymptotic balance is
\begin{eqnarray}
\Delta r j'\tilde{n}_m(t)\sim\frac{1}{a}\tilde{\psi}_m(t)\frac{\tilde{n}_m(t)}{\Delta r} &\hspace{1cm}&(j'\rightarrow\infty),
\end{eqnarray}
yielding
\begin{eqnarray}
\Delta r\sim\sqrt{\frac{\tilde{\psi}_m}{aj'}}&\hspace{1cm}&(j'\rightarrow\infty).
\label{delta}
\end{eqnarray}
The length $\Delta r$ is the scale of fluctuation variation within the coherent current filament.  In the simulations, the filaments were identified as regions of strong, localized, symmetric current surrounded by turbulent fluctuations.  Consequently, $\Delta r$ represents a fluctuation penetration depth into the structure.  We derived the layer width $\Delta r$ from linear and nonlinear kinetic Alfv\'{e}n wave terms operating on flux in the density equation.  Identical operators apply to $n$ in the flux equation.  Hence $\Delta r$ is the width of a single layer pertaining to both the density and current fluctuations of refracted KAW turbulence.  This structure is shown schematically in Fig.~2.  The above analysis indicates a single layer width and gives its value.  It does not give the functional variation of current and density fluctuations within the layer, either relative or absolute.

In the simpler case of intermittency in decaying 2D Navier-Stokes turbulence, statistical closure theory was used to derive spatial functions describing the inhomogeneity of turbulence in the presence of a coherent vortex \citep{ter92}.  There, coherent vortices suppress turbulent penetration via strong shear flow, analogous to role of refraction here.  For the KAW system the closure equations are much more complicated and not amenable to the WKB analysis that gave the functional form of the boundary layer in the Navier-Stokes case.  However, the closure remains useful.  It provides a mathematical platform from which to calculate all aspects of the interaction of filament and turbulence, including the accelerated decay of turbulence within the boundary layer, the spatial characteristics of the layer, and the amplitudes of $n$ and $\psi$.  These are necessary for calculating turbulent mixing rates of the filament current and density.

Closures can be applied to intermittent turbulence even though they rely  on Gaussian statistics.  The filaments, which make the system non Gaussian as a whole, are quasi stationary on the short time scale of turbulent evolution.  
Therefore, on that scale their only effect is to make the turbulence inhomogeneous.  The fast time scale statistics are a property of fast time scale nonlinearity, and remain Gaussian.  The closure equations are
\begin{equation}
\hat{\gamma}\tilde{\psi}_{m,\gamma}-im(r-r_0)j'\tilde{n}_{m,\gamma}-D_{\psi\psi}(m,\gamma)\frac{\partial^2}{\partial r^2}\nabla^2\tilde{\psi}_{m,\gamma}-D_{\psi n}(m,\gamma)\frac{\partial^2}{\partial r^2}\tilde{n}_{m,\gamma}=\frac{im}{r}\tilde{\psi}_{m,\gamma}\frac{d}{dr}n_0(r),
\end{equation}
\begin{eqnarray}
\hat{\gamma}\tilde{n}_{m,\gamma}&-&im(r-r_0)j'\nabla^2\tilde{\psi}_{m,\gamma}-D_{n\psi}^{(1)}(m,\gamma)\frac{\partial^2}{\partial r^2}\nabla^2\tilde{\psi}_{m,\gamma}-D_{n\psi}^{(2)}(m,\gamma)\frac{\partial^2}{\partial r^2}\nabla^4\tilde{\psi}_{m,\gamma} \nonumber \\ [2mm]
&-&D_{nn}^{(1)}(m,\gamma)\frac{\partial^2}{\partial r^2}\nabla^2\tilde{n}_{m,\gamma}-D_{nn}^{(2)}(m,\gamma)\frac{\partial^2}{\partial r^2}\tilde{n}_{m,\gamma}=\frac{im}{r}\tilde{\psi}_{m,\gamma}\frac{d}{dr}J_0(r).
\end{eqnarray}
This system is complex.  The six diffusivities all contribute to the lowest order as $j'\rightarrow \infty$.   (The diffusion coefficients and derivatives are not of the same order, but their product is.)   Moreover there is varied dependence on fluctuation correlations, and there are complex turbulent decorrelation functions.  For example
\begin{eqnarray}
D_{\psi\psi}(m,\gamma)&=&\frac{1}{2\pi i}\int_{-i\infty+\gamma_0}^{i\infty+\gamma_0}d\gamma'\sum_{m'\neq0,m}\frac{m'}{r}\Bigg\{\Big\langle\tilde\psi_{m',\gamma'} P_{m-m'}^{-1}\Delta W_{\gamma,\gamma'}\frac{m'}{r}{\psi}_{-m',-\gamma'}\Big\rangle \nonumber \\
&+&K_1(m-m',\gamma-\gamma')P_{m-m'}^{-1}\Delta W_{\gamma,\gamma'}\Big\langle\psi_{-m',-\gamma'}\tilde{n}_{m',\gamma'}\Big\rangle\Bigg\},
\end{eqnarray}
where
\begin{eqnarray}
P_{m}&=&\Bigg\{-\Bigg[-imj'(r-r_0)\nabla_{m}^2-D_{n\psi}^{(1)}(m,\gamma)\frac{\partial^2}{\partial r^2}\nabla_{m}^2-D_{n\psi}^{(2)}(m,\gamma)\frac{\partial^2}{\partial r^2}\nabla_{m}^4-\frac{im}{r}\frac{d}{dr}J_0\Bigg]K_1(m,\gamma) \nonumber \\ [2mm]
&+&\hat{\gamma}-D_{nn}^{(1)}(m,\gamma)\frac{\partial^2}{\partial r^2}\nabla_{m}^2-D_{nn}^{(2)}(m,\gamma)\frac{\partial^2}{\partial r^2}\Bigg\},
\end{eqnarray}
\begin{equation}
K_1(m,\gamma)=\Big(-\hat{\gamma}+D_{\psi\psi}(m,\gamma)\frac{\partial^2}{\partial r^2}\nabla_{m}^2-\frac{im}{r}\frac{d}{dr}n_0\Big)^{-1}\Big(imj'(r-r_0)+D_{\psi n}(m,\gamma)\frac{\partial^2}{\partial r^2}\Big),
\end{equation}
and $\Delta W(\gamma,\gamma')$ is the decorrelation rate for fluctuations at $\gamma'$ driving $\gamma$. 
Expressions for the other diffusivities are given in the appendix.  We now calculate turbulence properties from the closure equations.

\subsection{Decay Time for Turbulence in the Filament}

The time scale of turbulent evolution in the filament is given by $\hat{\gamma}$.  As $j'\rightarrow \infty$ this is dominated by the refraction.  Hence the first terms of Eq.~(23) and Eq.~(24) must balance the second terms, which in turn, must balance the remaining nonlinear terms.  If Eqs.~(23) and (24) are solved jointly retaining the first two terms, $\hat{\gamma}\sim im(r-r_0)j'\nabla$ as $j'\rightarrow \infty$.  Because $(r-r_0)\sim\Delta r$ and $\nabla\sim 1/\Delta r$,
\begin{eqnarray}
\hat{\gamma}\sim imj'&\hspace{1cm}&(j'\rightarrow\infty).
\end{eqnarray}
This time scale is purely imaginary, {\it i.e.}, oscillatory, when derived from a balance with only the linear Alfv\'{e}n terms.  When the diffusivities are included, it is complex.  This rapid decay suppresses turbulence in the filament relative to levels outside the filament.

\subsection{Alfv\'{e}nic Boundary Layer Width}

The width $\Delta r$, as derived previously, comes from independent balances in the equations for $n$ and $\psi$, and does not account for the kinetic Alfv\'{e}n wave dynamics that links $n$ and $\psi$.  To do that, Eqs.~(23) and (24) are combined into a single equation by operating on Eq.~(24) with $-im(r-r_0)j'-D_{\psi n}(m,\gamma)\partial^2/\partial r^2$ and substituting from Eq.~(23).  The resulting equation is eighth order in the radial derivative, and unsuitable for WKB analysis.  However, we can determine the radial scale as $j'\rightarrow \infty$ by dimensional anlaysis, taking $\nabla^2\sim \partial^2/\partial r^2 \sim 1/\Delta r^2$ and solving algebraically.  This is the same procedure used to obtain Eq.~(\ref{delta}).  Formally treating $\Delta r$ as a small parameter, we account for the fact that the diffusion coefficients have different scalings with respect to $\Delta r$, based on different numbers of radial derivatives operating on quantities within the coefficients.  Arbitrarily taking $D_{nn}^{(2)}$ as a reference diffusion coefficient, the definitions in the appendix show that if we define $D_{n\psi}^{(1)}=\Delta r d_{n\psi}^{(1)}$, $D_{n\psi}^{(2)}=(\Delta r)^3 d_{n\psi}^{(2)}$, $D_{nn}^{(1)}=(\Delta r)^2 d_{nn}^{(1)}$, $D_{nn}^{(2)}=d_{nn}^{(2)}$, $D_{\psi\psi}=(\Delta r)^2 d_{\psi\psi}$, and $D_{\psi n}=\Delta r d_{\psi n}$, then the lower case diffusivities $d_{n\psi}^{(1)}$,  $d_{n\psi}^{(2)}$, $d_{nn}^{(1)}$, $d_{nn}^{(2)}$, $d_{\psi\psi}$, and $d_{\psi n}$ are all of the same order.  We formally order the large parameter $j'$ by taking $mj'\rightarrow mj'/\epsilon$ and $\hat{\gamma}\rightarrow \hat{\gamma}/\epsilon$, where the controlling asymptotic limit becomes $\epsilon \rightarrow 0$.  The relationship between $\epsilon$ and $\Delta r$ will be derived by requiring that the asymptotic balance be consistent.  After all leading order expressions are derived, $\epsilon$ is set equal to 1.  

Substituting these relations into Eqs.~(23) and (24) and solving, we obtain:
\begin{eqnarray}
\Bigg\{\frac{\epsilon^4}{\Delta r^4}\Big[d_{\psi\psi}\Big(d_{nn}^{(1)}+d_{nn}^{(2)}\Big)-d_{\psi n}\Big(d_{n \psi}^{(1)}+d_{n \psi}^{(2)}\Big)\Big]&-&\frac{\epsilon^2}{\Delta r^2}\Big[imj'\Big(d_{n \psi}^{(1)}+d_{n \psi}^{(2)}+d_{\psi n}\Big)-\gamma\Big(d_{nn}^{(1)}+d_{nn}^{(2)} \nonumber \\ [2mm]
-d_{\psi\psi}\Big)\Big]+m^2j'^2+\hat{\gamma}^2\Bigg\}\tilde{\psi}_{m,\gamma}^{(i)}=\Bigg[\hat{\gamma}&-&\frac{d_{nn}^{(1)}+d_{nn}^{(2)}}{\Delta r^2}\Bigg]S_{\psi}+\Big[imj'+d_{\psi n}\Big]\Delta r S_n,
\end{eqnarray}
where 
\begin{eqnarray}
S_n&=&\frac{im}{r_0}\tilde{\psi}_{m,\gamma}\frac{d}{dr}J_0(r) \\
S_{\psi}&=&\frac{im}{r_0}\tilde{\psi}_{m,\gamma}\frac{d}{dr}n_0(r).
\end{eqnarray}
are the turbulence sources described in the previous section.  The left hand side is a dimensional representation of a Green function operator that governs the response to the sources.  
The spatial response decays inward from the edge of the filament where both the sources and the shear in $B_{\theta}$ are strong.  Consequently, the field $\tilde{\psi}_{m,\gamma}$ appearing in the sources $S_n$ and $S_{\psi}$ is understood to be characteristic of the filament edge, and therefore ambient turbulence, while $\tilde{\psi}_{m,\gamma}^{(i)}$ is a response accounting for the the refractive decay inside the filament.
The scale length of the response $\Delta r$ is found by solving the homogeneous problem, {\it i.e.}, by setting the left hand side equal to zero and solving for $\Delta r$.  
In the limit $\epsilon \rightarrow 0$, turbulence remains in the dynamics and contributes to $\Delta r$ only if $\epsilon \sim \Delta r$.  Otherwise, the dynamics is laminar.  The solution for $\Delta r$ is 
\begin{equation}
\Big(\frac{\epsilon}{\Delta r}\Big)^2=\frac{(imj'\hat{d}_2-\hat{\gamma}\hat{d}_3)}{2\hat{d}_1^2}+\frac{1}{2\hat{d}_1^2}\Bigg[\Big(imj'\hat{d}_2-\hat{\gamma}\hat{d}_3\Big)^2-4\hat{d}_1^2\Big(m^2j'^2+\hat{\gamma}^2\Big)\Bigg]^{1/2}
\end{equation}
where $\hat{d}_1^2=d_{\psi\psi}(d_{nn}^{(1)}+d_{nn}^{(2)})-d_{\psi n}(d_{n\psi}^{(1)}+d_{n\psi}^{(2)})$, $\hat{d}_2=d_{n\psi}^{(1)}+d_{n\psi}^{(2)}+d_{\psi n}$, and $\hat{d}_3=d_{nn}^{(1)}+d_{nn}^{(2)}-d_{\psi\psi}$.
This is the Alfv\'{e}nic generalization of Eq.~(22).  It is more complicated but gives identical scaling.  Recalling that all the lower case diffusivities have the same scaling and replacing them with a generic $d$, the solution scales as $\epsilon^2/\Delta r^2 \sim mj'/d$.  Setting $\epsilon=1$, 
\begin{eqnarray}
\Delta r \sim \sqrt{\frac{d}{mj'}}&\hspace{1cm}&(j'\rightarrow\infty).
\end{eqnarray}
The generic diffusivity $d$ can be evaluated from the definitions given for specific diffusivities in the appendix.  If the turbulent decorrelation functions are evaluated in a strong turbulence regime (turbulence time scales $<<$ linear time scales),
$d \sim \tilde{\psi}_m/a$, reproducing Eq.~(22).  

\subsection{Boundary Layer Structure of Turbulence}

Although the structure function has not been solved (just its radial scale), its form in simpler cases illustrates the rapid decay of turbulence across the boundary layer, from the edge inward.   Where WKB analysis is possible, the leading order spatial Green function has the form
\begin{eqnarray}
G(r|r') \sim \textrm{exp}\Bigg\{-C\Bigg(\frac{r_{<}-r_0}{\sqrt{d/mj'}}\Bigg)^\alpha\Bigg\}\textrm{exp}\Bigg\{C\Bigg(\frac{r_{>}-r_0}{\sqrt{d/mj'}}\Bigg)^\alpha\Bigg\}, 
\end{eqnarray}

\noindent{where} $C$ is a complex constant with positive real part, $r_{<}$ ($r_{>}$) is the smaller (larger) of $r$ and $r'$, and $\alpha$ is a positive constant determined by the order of the homogeneous operator. 
Here our dimensional solution of the problem, carried out by inverting Eq.~(29), captures the radial integral over a structure function like that of Eq.~(34).   
Solving Eq.~(29) we obtain
\begin{equation}
\tilde{\psi}_{m,\gamma}^{(i)} \sim \hat{\gamma}^{-2}\Bigg\{ \Bigg[\hat{\gamma}-\frac{d_{nn}^{(1)}+d_{nn}^{(2)}}{\Delta r^2}\Bigg]S_{\psi}+\Big[imj'+d_{\psi n}\Big]\Delta r S_n \Bigg\} \sim \hat{\gamma}^{-1}\frac{m}{a}\hat{\psi}(r_0)[n_0'+\Delta r J_0'].
\end{equation}
The temporal and spatial response to turbulent sources $S_{\psi}$ and $S_n$ at a point $r_0$ in the filament edge appears here as a structure factor of magnitude $\hat{\gamma}^{-2}$ multiplying the source.  The product of source and response yields the value of $\tilde{\psi}^{(i)}$ inside the boundary layer.  Beyond $\Delta r$ the response decays with an envelope like that of Eq.~(34).
The part of the source proportional to $n_0'$ is essentially larger than the part proportional to $J_0'$ by O($a/\Delta r$).  However if $\tilde{\psi}^{(i)}$ is substituted into the correlations of the equation for $\psi_0$ [Eq.~(12)], the $J_0'$-part yields the diagonal terms.
The density is given by the dimensional representation of Eq.~(23),
\begin{equation}
\tilde{n}_{m,\gamma}^{(i)}\sim-\Big[imj'\Delta r+\frac{d_{\psi n}}{\Delta r}\Big]^{-1}\Big[S_{\psi}+\frac{D_{\psi\psi}}{\partial r^2}\tilde{\psi}_{m,\gamma}-\hat{\gamma}\tilde{\psi}_{m,\gamma}\Big]\sim\frac{\tilde{\psi}_{m,\gamma}(r_0)\Big[n_0'+\Delta r J_0'\Big]}{aj'\Delta r}.
\end{equation}

\subsection{Condition for Strong Refraction}

The layer width $\Delta r$ is both the embodiment of the strong refraction of turbulent KAW activity in the filament by the large magnetic field shear $j'$, and a condition for the refraction to be sufficiently strong to modify the scales of turbulence in the filament relative to those outside it.  With $a$ the scale of typical fluctuations of interest, the refraction is strong when $\Delta r/a\ll1$, or 
\begin{equation}
\frac{\Delta r}{a}\sim\sqrt{\frac{d|_{r>a}}{a^2mj'}} \textrm{    }\sim \textrm{    } \sqrt{\frac{\tilde{\psi}|_{r>a}}{a^3mj'}} \textrm{   } \ll \textrm{   } 1.
\end{equation}
As a condition for strong refraction it makes sense to use values for $d$ or $\psi$ that are typical of the turbulence in regions $r>a$ where there are no intense filaments.  Inside a strong filament the reduction of turbulent KAW activity represented by the structure factor $\hat{\gamma}^{-2}$ makes $d|_{r<a}\ll d|_{r>a}$.  Accordingly, the boundary layer width $\Delta r/a$ is smaller than $[d|_{r>a}/a^2mj']^{1/2}$.

\section{Filament Decay from Mixing Stresses}

The long time evolution of the filament fields $\psi_0(\tau)$ and $n_0(\tau)$ is governed by the mixing stresses of Eqs.~(12) and (13).  These can now be evaluated using the boundary layer responses $\tilde{\psi}_{m,\gamma}^{(i)}$ and $\tilde{n}_{m,\gamma}^{(i)}$ derived in the previous section.  Because these fields are confined to the layer, the time scale $\tau$ is a mixing time across the boundary layer.  For the diagonal stress components, the mixing is diffusive.  The asymptotic behavior of the boundary layer yields the following dimensional equivalents: $r-r_0\sim\Delta r$ and $\partial/\partial r \sim 1/\Delta r$, as before; $\int dr \sim \Delta r$; $\int d\gamma \sim mj'$; $\tilde{\psi}_{m,\gamma} \sim \psi_m(t)/mj'|_{t=0}$; and $\int d\gamma' \langle\tilde{\psi}\tilde{n}\rangle \sim \psi(t)n(t)/mj'|_{t=0}$. (The latter two expressions are inverse Laplace transform relations.)  With these conventions
\begin{eqnarray}
\frac{\partial\psi_{\gamma=0}(\tau)}{\partial \tau}\approx\frac{\psi_{\gamma=0}}{\tau_{\psi}}&=&-\frac{1}{2\pi i}\int_{-i\infty+\gamma_0}^{-i\infty+\gamma_0}d\gamma'\sum_{m'}\Bigg[\frac{im'}{r}\Big(\Big\langle\tilde{\psi}_{-m',-\gamma'}^{(i)}\frac{\partial }{\partial r}\tilde{n}_{m',\gamma'}\Big\rangle+\Big\langle\tilde{\psi}_{-m',-\gamma'}\frac{\partial }{\partial r}\tilde{n}_{m',\gamma'}^{(i)}\Big\rangle\Big) \nonumber \\
&+&\Big(\Big\langle\frac{\partial}{\partial r}\tilde{\psi}_{-m',-\gamma'}^{(i)}\tilde{n}_{m',\gamma'}\Big\rangle+\Big\langle\frac{\partial}{\partial r}\tilde{\psi}_{-m',-\gamma'}\tilde{n}_{m',\gamma'}^{(i)}\Big\rangle\Big)\frac{im'}{r}\Bigg] \nonumber \\
&\approx&\sum_m\frac{1}{a^2j'^2}\Big[\langle\tilde{b}_{\theta -m}\tilde{n}_m\rangle |_{t=0}+\langle\tilde{b}_{\theta m}^2\rangle|_{t=0}\Big]\Big[n_0'+\Delta rJ_0'\Big],
\end{eqnarray}
where $\tilde{b}_{\theta m}=\tilde{\psi}_m/\Delta r$.  The factor $j'^{-2}$ in the right-most form makes $\tau_{\psi}$ large, {\it i.e.}, mixing across the boundary layer is impeded by refraction.  The turbulent fields in these expressions are filament edge fields, {\it i.e.}, they are characteristic of ambient turbulence.  The mixing time for current can be obtained by operating with $\nabla^2$ on both sides of the Eq.~(38).  On the left hand side $\nabla^2\psi_{\gamma=0}\rightarrow\psi_{\gamma=0}/a^2$, while on the right hand side, $\nabla^2\rightarrow1/\Delta r^2$.  Consequently, 
\begin{equation}
\frac{\psi_{\gamma=0}}{\tau_J}\approx\sum_m\frac{1}{\Delta r^2j'^2}\Big[\langle\tilde{b}_{\theta -m}\tilde{n}_m\rangle |_{t=0}+\langle\tilde{b}_{\theta m}^2\rangle|_{t=0}\Big]\Big[n_0'+\Delta rJ_0'\Big].
\label{current0it}
\end{equation}
This time scale is much shorter because current, as a second derivative of $\psi$, has finer scale structure.  If the filament is Alfv\'{e}nic, {\it i.e.}, $n_0\approx aJ_0$, the mixing time is dominated by the part of Eq.~(39) that is proportional to $n_0'$.  This represents off-diagonal transport of current driven by density gradient.  The diagonal transport (driven by $J_0'$) is current diffusion, and is slower by a factor $a/\Delta r$.  In the discussions that follow we will deal with the current diffusion time scale, although similar behavior will hold for the off diagonal transport.
The mixing time for density is 
\begin{equation}
\frac{\partial n_{\gamma=0}(\tau)}{\partial \tau}\approx\frac{n_{\gamma=0}}{\tau_{n}}\approx\sum_m\frac{2\langle\tilde{b}_{\theta m}^2\rangle|_{t=0}}{a^2j'^2}\Big(\frac{n_0'}{\Delta r}+J_0'\Big).
\end{equation}
Here the dominant component (proportional to $n_0'$) is diffusive. 

We evaluate these boundary layer mixing times relative to the two turbulent time scales of the system.  These are $\gamma^{-1}$, the turbulent decay time in the layer, and $\tau_A=a^2/\tilde{b}_{\theta}|_{r\gg a}$, a turbulent Alfv\'{e}n time outside the filament.  To simplify expressions we note that Alfv\'{e}nic equipartition implies that $\langle \tilde{b}_{\theta}\tilde{n}\rangle\approx\langle \tilde{b}_{\theta}^2\rangle$.  We also note that $\psi_{\gamma=0}$ is in the Laplace transform domain, whereas $n_0'$ and $J_0'$ are in the time domain.  Under the inverse Laplace transform, $\psi_{\gamma=0}\gamma\approx\psi(t,\tau)\equiv\psi_0$.  The scale of the filament is $a$, so $\psi_{\gamma=0}\gamma\approx a^2J_0$.   Similarly, $n_0'\approx n_0/a$, $J_0'\approx J_0/a$, and $j'=\partial/\partial r(B_{\theta}/r)\approx B_{\theta}/a^2$.  We assume the filament is Alfv\'{e}nic, making $n_0\approx aJ_0$.  With these relations, 
\begin{equation}
\tau_n\gamma=\tau_{\nabla^2\psi}\gamma=\frac{\Delta r}{a}\frac{B_{\theta}^2}{\tilde{b}_{\theta}^2}\sim\Big(\frac{a}{\Delta r}\Big)^3\sim\Big(\frac{B_{\theta}}{\tilde{b}_{\theta}}\Big)^{3/2}.
\end{equation}
The last two equalities make use of Eq.~(37), and the fact that $\Delta r/a$ and the mixing fluctuations are referenced to ambient turbulence levels for which $\tilde{b}_{\theta}\sim\tilde{\psi}/a\sim\tilde{b}_r$.  Equation (41) indicates that turbulent diffusion times across the mixing layer $\Delta r$ for both $n_0$ and $J_0$ are comparable, and are much longer than the decay times of turbulence in the layer.  The strong shear limit, previously indicated by $j'\rightarrow\infty$, is here replaced by $B_{\theta}\rightarrow\infty$, because with a fixed radius $a$, strong shear means large $B_{\theta}$.  In terms of $\tau_A$,
\begin{equation}
\tau_n/\tau_A=\tau_{\nabla^2\psi}/\tau_A=\frac{\Delta r}{a}\frac{B_{\theta}}{\tilde{b}_{\theta}}\sim\Big(\frac{a}{\Delta r}\Big)\sim\Big(\frac{B_{\theta}}{\tilde{b}_{\theta}}\Big)^{1/2},
\end{equation}
indicating that these diffusion times are longer than the Alfv\'{e}nic time of the ambient turbulence.

Either of the above expressions indicates that the actual lifetime of a filament (as opposed to the turbulent diffusion time across the edge layer) is virtually unbounded, provided direct damping due to resistivity or collisional diffusion is negligible.  During a filament lifetime turbulence must diffuse across the scale $a$, many $\Delta r$-layer widths from the filament edge of to its center.  However, in just a layer time $\tau_n$ or $\tau_{\nabla^2\psi}$, the turbulence is reduced by many factors of $e^{-1}$, while the filament density or current inside of the layer remains untouched.  Consequently the width of the mixing layer at the edge of the filament continuously decreases even as the time to mix across it increases.  The result is that mixing never extends to the filament core.  This analysis shows that structures identified in the simulations as current filaments correlate spatially with a coherent density field, provided the density component is not destroyed by strong collisional diffusion.

\section{Geometric and Statistical Properties}

The above analysis treats the current of the filament as localized.  The current is maximum at $r=0$ and becomes zero at $r=a$.  This makes the shear of the filament magnetic field largest in the filament edge, and zero in the center.  If it is true that the shear of this field refracts turbulent KAW activity as described above, turbulence is suppressed where the shear is large.  These properties are incorporated in the spatial variation of a single quantity known as the Gaussian curvature \citep{ter00}.  The Gaussian curvature is a property of vector fields that quantifies the difference between shear stresses and rotational behavior.   In rectilinear coordinates the Gaussian curvature $C_T$ of a vector field {\bf A}$(x,y)$ is \citep{mcw84}
\begin{equation}
C_T=\Big[\frac{\partial A_x}{\partial x}-\frac{\partial A_y}{\partial y}\Big]^2+\Big[\frac{\partial A_y}{\partial x}+\frac{\partial A_x}{\partial y}\Big]^2-\Big[\frac{\partial A_y}{\partial x}-\frac{\partial A_x}{\partial y}\Big]^2.
\end{equation}
For the total magnetic field in our cylindrical system this can be written
\begin{equation}
C_T=\Big[r\frac{d}{dr}\Big(\frac{\tilde{b}_{r}}{r}\Big)-\frac{1}{r}\frac{\partial}{\partial \theta}\tilde{b}_{\theta}\Big]^2+\Big[r\frac{d}{dr}\Big(\frac{B_{\theta}+\tilde{b}_{\theta}}{r}\Big)+\frac{1}{r}\frac{\partial}{\partial \theta}\tilde{b}_r\Big]^2-\Big[J_0+\tilde{j}\Big]^2.
\end{equation}
Inside the filament, turbulence is suppressed, and $C_T$ is dominated by the filament field components $B_{\theta}$ and $J_0$.  Near the center, $J_0$ is maximum and $d(B_{\theta}/r)/dr=j'$ vanishes, making $C_T$ negative.  Toward the filament edge, $j'$ becomes maximum as $J_0$ goes to zero, making $C_T$ positive.  Outside the filament $C_T$ is governed by $\tilde{b}_{\theta}$, $\tilde{b}_r$, and $\tilde{j}$.  These components must be roughly in balance.  If they are not, the conditions for forming a coherent filament are repeated, and a structure should be present.  Therefore, in regions where there are coherent filaments, the Gaussian curvature should have a strongly negative core surrounded by a strongly positive edge.  Where there are no coherent structures the Gaussian curvature should be small.   If this property is observed in simulations, it confirms the hypothesis that shear in the filament field refracts turbulent KAW activity in such a way as to suppress turbulent mixing of the structure.  We note that the negative-core/positive-edge structure is predicted for current filaments of either sign, positive or negative.  This type of Gaussian curvature structure has been observed in recent simulations \citep{smi06}.

If the current filaments are well separated, their slow evolution relative to the decaying turbulence that surrounds them leads to a highly non Gaussian PDF.  Assuming an initial PDF that is Gaussian with variance $\langle J_{\sigma}^2 \rangle$,
\begin{equation}
P(J)=\frac{1}{\sqrt{2\pi}\langle J_{\sigma}^2 \rangle^{1/2}}\exp\Big[\frac{-J^2}{2\langle J_{\sigma}^2 \rangle}\Big],
\end{equation}
it is possible to model subsequent evolution on the basis of the time scales derived previously and the condition for strong refraction, Eq.~(37).  This condition stipulates that structures form where refraction is large, {\it i.e.}, where $mj'\gg\tilde{\psi}|_{r>a}/a^3$.  Since $j'\approx J_0/a$ and $\tilde{\psi}|_{r>a}/a^2\approx  \langle J_{\sigma}^2 \rangle^{1/2}$, structures occur for $J_0\geq J_c=C\langle J_{\sigma}^2 \rangle^{1/2}$,  where $C$ is the smallest numerical factor above unity to guarantee strong refraction and mixing suppression.  Given the latter, filaments reside on the tail of the PDF with high $J$ and low probability $\int_{J_c}^{\infty} P(J)dJ$.  This probability is equal to the filament packing fraction, {\it i.e.}, the fraction of 2D space occupied by current filaments.  If, for simplicity, we assume that all filaments are of radius $a$, the faction of 2D space they occupy is $(a/l)^2$, where $l$ is the mean distance between filaments.  Therefore,
\begin{equation}
\int_{-\infty}^{-J_c} P(J)dJ+\int_{J_c}^{\infty} P(J)dJ=2\int_{J_c}^{\infty} P(J)dJ=\Big(\frac{a}{l}\Big)^2,
\end{equation}
where we assume that $P(J)$ is an even function.  This expression gives the packing fraction as a function of the critical current $J_c$ for filament formation.

It is now straightforward to construct a heuristic model for the evolution from the initial distribution.  The model applies for times that are larger than the turbulent Alfv\'{e}n time, but shorter than the mean time between filament mergers.  (Once filaments begin merging, their number and probability begin decreasing.)  Prior to that time the filament part of the distribution with $J>J_c$ is essentially unchanged, apart from the minor effects of slow erosion at the edge of the filaments.  The probability that a fluctuation is not a filament also remains fixed, but these fluctuations decay in time.  This means that the variance decreases while the probability remains fixed.  The rate of decay is the turbulent Alfv\'{e}n time $\tau_A$.  Therefore the distribution can be written
\begin{eqnarray}
P(J,t)=\frac{N(t)}{\sqrt{2\pi}\langle J_{\sigma}^2\rangle ^{1/2}}\exp\Bigg[\frac{-J^2}{2\langle J_{\sigma}^2\rangle\exp[-t/\tau_A]}\Bigg] &\hspace{1.0 cm}& \textrm{ for }J<J_c \nonumber \\ [0.2in]
P(J,t)=\frac{1}{\sqrt{2\pi}\langle J_{\sigma}^2\rangle ^{1/2}}\exp\Bigg[\frac{-J^2}{2\langle J_{\sigma}^2\rangle}\Bigg] &\hspace{1.8cm}& \textrm{ for }J\geq J_c
\end{eqnarray}
where $\langle J_{\sigma}^2\rangle$ remains the initial variance, and $N(t)$ is a time-dependent normalization constant that maintains $\int_0^{J_c}P(J,t)dJ$ at its initial value, {\it i.e.},
\begin{equation}
N(t)=\frac{\int_0^{J_c}dJ\exp\Big[-J^2/2\langle J_{\sigma}^2 \rangle\Big]}{\int_0^{J_c}dJ\exp\Big[-J^2\exp[t/\tau_A]/2\langle J_{\sigma}^2 \rangle\Big]}.
\end{equation}
The distribution $P(J,t)$ becomes highly non Gaussian as $t\gg\tau_A$ because one part of the distribution (for $J<J_c$ ) collapses onto the $J=0$ axis and becomes a delta function $\delta(J)$, while the other part remains fixed.

A simple measure of the deviation from a Gaussian distribution is the kurtosis, 
\begin{equation}
\kappa(t)=\frac{3\int J^4P(J,t)dJ }{\Big[\int J^2P(J,t) dJ\Big]^2}.
\end{equation}
The evolving kurtosis can be calculated directly from Eq.~(47).  While the exact expression is not difficult to obtain, its asymptote is more revealing.  The kurtosis diverges from the initial Gaussian value of 3 as the contribution from turbulent kinetic Alfv\'{e}n waves ($J<J_c$) decays and collapses to $\delta(J)$.  After a few Aflv\'{e}n times the kurtosis is dominated by the part with $J>J_c$, which, because it is stationary, represents the time asymptotic value for $\tau_A< t < \tau_{M}$.  The time $\tau_M$ is the mean time to the first filament mergers.  The time asymptotic kurtosis is 
\begin{equation}
\kappa(\tau_A\ll t\ll\tau_M)=\frac{6\int_{J_c}^{\infty}P(J)J^4dJ}{\Big[2\int_{J_c}^{\infty}P(J)J^2dJ\Big]^2}=\frac{3}{2}\Big(\frac{l}{a}\Big)^2\Big[1+\frac{\langle J_{\sigma}^2\rangle}{J_c^2}+\textrm{O}\Big(\frac{\langle J_{\sigma}^2\rangle^{3/2}}{J_c^3}\Big)\Big].
\end{equation}
In writing this expression, the left hand side of Eq.~(46) has been expanded for $J_c^2>\langle J_{\sigma}^2 \rangle$ to yield $\langle j_{\sigma}^2 \rangle\exp(-J_c^2/2\langle J_{\sigma}^2\rangle)=(a/l)^2J_c[1+\textrm{O}(\langle J_{\sigma}^2\rangle^{3/2}/J^3)]$.  The time-asymptotic kurtosis is much greater than the initial Gaussian value of 3 and is characterized by the initial value of the inverse packing fraction.   Once filament mergers begin, the inverse packing fraction increases above the initial value $(l/a)^2$.  If $[l(t)/a]^2$ is the inverse packing fraction for $t>\tau_M$, the above analysis suggests that the kurtosis will continue increasing as $(3/2)[l(t)/a]^2$ for late times. 

We now consider the distribution of density.  As shown in the previous section, the density present in the current filament also has suppressed mixing and is therefore coherent, or long lived.  However, it is not spatially intermittent to the same degree as the current.  Alfv\'{e}nic dynamics indicate that $n\approx B$, while Ampere's law stipulates that the magnetic field of the filament extends into the region $r>a$, falling off as $r^{-1}$.  Hence the density associated with filaments also is expected to fall off as $r^{-1}$ for $r>a$.  This spatially extended structure makes density less isolated than current.  It produces higher probabilities for low values of density than those of decaying turbulence.  This will yield a kurtosis closer to the Gaussian value of 3 than the kurtosis of the current.  However, the distribution of low-level density associated with the structure likely will not be Gaussian.  It is ultimately the distribution that matters for the scattering of {\it rf}-pulsar signals.  

To construct the density PDF we seek the mapping of density onto the spatial area it occupies.  We obtain this mapping for the filament density, assuming that the density of turbulence is low, and has effectively collapsed onto $n=0$ after a few $\tau_A$, just as the current.  The density for $r>a$ goes as $n=an_0/r$, where $n_0$ is the value of the density at $r=a$.  As shown in Fig. 3, the area occupied for a given density is $2\pi rdr$.  This area is the probability when properly normalized, hence, 
\begin{equation}
P(n)dn=2\pi rdr
\end{equation}
Writing $rdr$ in terms of $ndn$ using $n=an_0/r$, 
$P(n)=C_n/n^3$, where $C_n$ is the normalization constant chosen so that the probability integrated over the whole filament with its $r^{-1}$-mantle equals the packing fraction, or probability of finding the filament in some sample area.
With the long, slowly decaying tail of $n^{-3}$ it is necessary to impose a cutoff to keep the PDF integrable.  The cutoff, which we will label $n_c$, corresponds to the low level of decaying turbulence, but otherwise need not be specified.  Consequently, the normalization is determined by
\begin{eqnarray}
2C_n\int_{n_c}^{n_0}\frac{dn}{n^3}&=&\big(\frac{r_c}{l}\big)^2\hspace{1cm}\textrm{for }r_c<l, \nonumber \\ [0.2in]
2C_n\int_{n_c}^{n_0}\frac{dn}{n^3}&=&1 \hspace{1.7cm}
\textrm{for }r_c\geq l ,
\end{eqnarray}
where $r_c$, the radius at which $n=n_c$, is $r_c=an_0/n_c$.  The first of the two possibilities in Eq.~(52) allows for a cutoff radius that is smaller than the mean distance between structures (of radius $r_c$), yielding a probability that is less than unity.  If the cutoff radius is equal to or larger than the mean separation, then the structures are space filling and the probability is unity.  Solving for $C_n$, the normalized density PDF is
\begin{eqnarray}
P(n)=\frac{a^2}{l^2}\frac{n_0^4}{(n_0^2-n_c^2)n^3}&\hspace{1.0cm}&r_c<l\hspace{0.5cm}\textrm{or}\hspace{0.5cm}\frac{a}{l}<\frac{n_c}{n_0} \nonumber \\ [0.2in]
P(n)=\frac{n_0^2n_c^2}{(n_0^2-n_c^2)n^3} &\hspace{1.2cm}&r_c\geq l\hspace{0.5cm}\textrm{or}\hspace{0.5cm}\frac{a}{l}\geq\frac{n_c}{n_0}
\end{eqnarray}
This distribution is defined for $n_c<n<n_0$.  It captures only the contribution of filaments, and ignores the density inside $r=a$, which makes a small contribution to the PDF. 

This distribution is certainly non Gaussian, because it has a tail that decays slowly as $n^{-3}$.  However, depending on the length of the tail, which is set by $n_c$ and $n_0$, the distribution may or may not deviate from a Gaussian is a significant way over $n_c<n<n_0$.  This is quantified by the kurtosis,
\begin{equation}
\kappa(n_0,n_c)=\frac{6C_n \int_{n_c}^{n_0}n^4(n^{-3}dn)}{C_n^2\Big[2\int_{n_c}^{n_0}n^2(n^{-3}dn)\Big]^2}.
\end{equation}
Substituting from Eq.~(53), we find that
\begin{eqnarray}
\kappa(n_0,n_c)=\frac{3}{4}\frac{l^2}{a^2}\frac{\Big[1-2n_c^2/n_0^2+n_c^4/n_0^4\Big]}{\Big[\ln\big(n_0/n_c\big)\Big]^2}, &\hspace{1cm}& r_c<l\hspace{.5 cm}\textrm{or}\hspace{.5 cm}\frac{a}{l}<\frac{n_c}{n_0},  \\ [0.2in]
\kappa(n_0,n_c)=\frac{3}{4}\frac{n_0^2}{n_c^2}\frac{\Big[1-2n_c^2/n_0^2+n_c^4/n_0^4\Big]}{\Big[\ln\big(n_0/n_c\big)\Big]^2}, &\hspace{1cm}& r_c\geq l\hspace{.5 cm}\textrm{or}\hspace{.5 cm}\frac{a}{l}\geq\frac{n_c}{n_0}.
\end{eqnarray}
These expressions are smaller than the current kurtosis by a factor $2[\ln(n_0/n_c)]^2$.  Unless $n_0/n_c$ is quite large, the kurtosis may not rise significantly above 3.  This is particularly true in simulations with limited resolution where dissipation will affect the density, either directly through a collisional diffusion, or indirectly by resistive diffusion of current filaments.  Kurtosis increases if $n_c$ decreases.  However, while $n_c$ is tied to the decreasing turbulence level, regeneration of the turbulence by the $r^{-1}$ mantle may prevent $n_c$ from becoming very small.  Nonetheless, mergers of filaments will decrease the packing fraction.  Even if the density is space filling initially and satisfies Eq.~(56), the mean filament separation will increase above $r_c$ at some point, and the kurtosis will be given by Eq.~(55).  Then as the inverse packing fraction increases above $l^2/a^2$, the kurtosis will rise. 

The $n^{-3}$ falloff of the density PDF has intriguing implications for {\it rf} scattering of pulsar signals.  Noting that the scattering is produced by gradients of density, the extended density structure for $r>a$ yields $\nabla n\equiv n' \sim 1/r^2$.  We can construct the PDF for $n'$ following the procedure used for the PDF of $n$.  Writing $rdr$ in terms of $n'dn'$ using $r \sim (n')^{-1/2}$, we recover
\begin{equation}
P(n')=\frac{C_{n'}}{(n')^2},
\end{equation}
where $C_{n'}$ is a constant.  This is a Levy distribution, the type of distribution inferred in the scaling of pulsar signals \citep{stas03b}.  Further exploration of the implications of these results to {\it rf} scattering of pulsar signals remains an important question for future work. 

\section{Conclusions}

We have examined the formation of coherent structures in decaying kinetic Alfv\'{e}n wave turbulence to determine if there is a dynamical mechanism in interstellar turbulence that leads to a non Gaussian PDF in the electron density.  Such a PDF has been inferred from scalings in pulsar scintillation measurements.  We use a model for kinetic Alfv\'{e}n wave turbulence that is applicable when there is a strong mean magnetic field.  The nonlinearities couple density and magnetic field in the plane perpendicular the mean field in a way that is analogous to the coupling of flow and magnetic field in reduced MHD.  The model applies at scales on the order of the ion gyroradius and smaller.  We show that the coherent current filaments previously observed to emerge from a Gaussian distribution in simulations of this model \citep{crad91} result from strong refraction of turbulent kinetic Alfv\'{e}n waves.  The refraction occurs in the edge of  intense, localized current fluctuations, and is caused by the strongly sheared magnetic field associated with the current.  This refraction localizes turbulent wave activity to the extreme edge of the filament, and impedes mixing (turbulent diffusion) of the filament current by the turbulence.  From this analysis we conclude that the turbulence suppression by sheared flows common in fusion plasmas \citep{ter00} has a magnetic analog in situations where there is no flow.  This leads to a further conclusion that intermittent turbulence, which is generally associated with flows, can occur in situations where there is no flow.  (By flow we mean ion motion.  Electron motion is incorporated in the current.)  We have derived a condition for the strength of magnetic shear required to produce the strong refraction and suppress mixing.  We show that this condition yields a prediction for the Gaussian curvature of the magnetic field.  This quantity is predicted to have large values inside the coherent current filaments, and small values everywhere else.  Inside filaments the Gaussian curvature is negative at the center, and positive at the edge.   

The analysis shows that long-lived fluctuation structures form in the density and magnetic field, provided damping is negligible.  Like the current filaments, these structures avoid mixing because of the refraction of turbulent kinetic Alfv\'{e}n wave activity.  Hence they occur in the same physical location as the current filaments.  However the localized nature of the current filaments give the long-lived magnetic field an extended region external to the current.  In this region the field falls off as $r^{-1}$, where $r$ is distance from the center of the filament.  Because kinetic Alfv\'{e}n wave dynamics yields an equipartition of density and magnetic field fluctuations, we posit that the long-lived density has a similar extended structure.  As a result, the connection between coherent structure and localization that is true for the current, and makes it highly intermittent, does not apply to the density.  While there is coherent long-lived density, it need not be localized.  A similar situation holds for vorticity and flow in 2D Navier Stokes turbulence \citep{mcw84}.  To explore this matter we have used the physics of the coherent structure formation to derive heuristic probability distribution functions for the current and density.  As the turbulence decays, leaving intense current fluctuations as coherent current filaments, the kurtosis of current increases to a value proportional to the packing fraction.  The kurtosis of density does not become as large, and could, under appropriate circumstances, remain close to the Gaussian value of 3.  However mergers of structures in a situation with very weak dissipation could increase the kurtosis well above 3.  More importantly, however, the density PDF is non Gaussian even when its kurtosis is not greatly different from 3.  The $r^{-1}$ structure external to the current gives the PDF a tail that goes as $n^{-3}$.  Mapping the $r^{-1}$ structure to a PDF in density gradient, the density-gradient PDF decays as $1/n'^2$, a Levy distribution.  This suggests that the mechanism described here may play a role in the scaling of pulsar {\it rf} signals.

Several aspects of this problem need additional study.  It is important to adapt these results to a steady state.  Generally speaking, there is a dynamical link between decaying turbulence and turbulence in a stationary dissipation range.  Hence these results, at least qualitatively, are relevant to the dissipation range.  Dissipation begins to affect the spectrum at a scale that is somewhat larger (order of magnitude) than the nominal dissipation scale \citep{fri95}.  Structures such as these would correspond to active, filamentary regions of dissipation analogous to those observed in neutral gas clouds, albeit at a much smaller scale and with no accompanying flow shear signature.  Intermittent structures can extend into the stationary inertial range, but the analysis presented here must be modified.  In the inertial range, turbulence is replenished, allowing the slow mixing of a coherent structure to continue until it is gone.  Structures are also regenerated by the turbulence, and the statistics is ultimately set by a balance of mixing and regeneration rates.  The mixing rates calculated here are sufficiently slow in strong filaments, that coherent structure formation is expected even in a steady state.  There is also a possible link between structures in the larger scale range of shear Alv\'{e}n excitations and KAW excitations.  These questions will be explored in future work.  While gyroradius-scale KAW turbulence may arise in astrophysical contexts other than the ISM, the small scales make it unlikely that astrophysical observations will be available for testing this theory. 
Therefore, simulations should be used to check key conclusions from the theoretical work presented here.  These include the formation of density structures, which was not reported in \citet{crad91}, the structure of the Gaussian curvature, which validates the refraction hypothesis, and the existence of the $r^{-1}$ structure in the density and its effect on the PDF.   The effect of this type of density field on {\it rf} scattering remains the underlying question, and modeling of the scattering with simulated fields should be pursued.

\acknowledgments

The authors acknowledge useful conversations with Stanislav Boldyrev, including his observation that the density PDF derived herein immediately leads to a Levy distribution in the density gradient.  PWT also acknowledges the Aspen Center for Physics, where part of this work was performed.  This work was supported by the National Science Foundation.

\appendix

\section{Appendix}

Closures truncate the moment hierarchy that is generated when averages are taken of nonlinear equations.  The closure we have used is of the eddy damped quasi normal Markovian variety, and follows the steps of the closure calculation described in \citep{ter01}.  The nonlinear decorrelation is calculated consistent with the statistical ansatz, not imposed ad hoc.  The closure equations are given in Eqs.~(23) and (24).  The other diffusivities not given in Eq.~(25) are
\begin{eqnarray}
D_{\psi n}(m,\gamma)&=&\frac{1}{2\pi i}\int_{-i\infty+\gamma_0}^{i\infty+\gamma_0}d\gamma'\sum_{m'\neq0,m}\frac{m'}{r}\Bigg\{\Big\langle\tilde\psi_{m',\gamma'}K_2(m-m',\gamma-\gamma')\Delta W_{\gamma,\gamma'}\frac{m'}{r}{\psi}_{-m',-\gamma'}\Big\rangle \nonumber \\
&+&\Big\langle\tilde{n}_{m',\gamma'}K_3(m-m',\gamma-\gamma'))\Delta W_{\gamma,\gamma'}\frac{m'}{r}{\psi}_{-m',-\gamma'}\Big\rangle\Bigg\},
\end{eqnarray}
\begin{equation}
D_{nn}^{(1)}(m,\gamma)=\frac{1}{2\pi i}\int_{-i\infty+\gamma_0}^{i\infty+\gamma_0}d\gamma'\sum_{m'\neq0,m}\frac{m'}{r}\Bigg\{\Big\langle\tilde\psi_{m',\gamma'}K_3(m-m',\gamma-\gamma')\Delta W_{\gamma,\gamma'}\frac{m'}{r}{\psi}_{-m',-\gamma'}\Big\rangle\Bigg\},
\end{equation}
\begin{eqnarray}
D_{nn}^{(2)}(m,\gamma)&=&\frac{1}{2\pi i}\int_{-i\infty+\gamma_0}^{i\infty+\gamma_0}d\gamma'\sum_{m'\neq0,m}\Bigg\{\frac{m'}{r}\Big\langle\tilde\psi_{m',\gamma'}\nabla^2K_3(m-m',\gamma-\gamma')\Delta W_{\gamma,\gamma'}\frac{m'}{r}{\psi}_{-m',-\gamma'}\Big\rangle \nonumber \\ [2mm]
&-&\frac{(m-m')}{r}\Big\langle\frac{\partial \tilde{\psi}_{m',\gamma'}}{\partial r}K_3(m-m',\gamma-\gamma')\Delta W_{\gamma,\gamma'}\Big(\frac{m}{r}\Big)\frac{\partial \tilde{\psi}_{-m',-\gamma'}}{\partial r}\Big\rangle \nonumber \\ [2mm]
&-&\frac{m}{r}\Big\langle K_3(m-m',\gamma-\gamma')\Delta W_{\gamma,\gamma'}\frac{m'}{r}\tilde{\psi}_{m',\gamma'}\nabla^2\tilde{\psi}_{-m',-\gamma'}\Big\rangle \Bigg\},
\end{eqnarray}
\begin{equation}
D_{n\psi}^{(2)}(m,\gamma)=\frac{1}{2\pi i}\int_{-i\infty+\gamma_0}^{i\infty+\gamma_0}d\gamma'\sum_{m'\neq0,m}\frac{m'}{r}\Bigg\{\Big\langle\tilde\psi_{m',\gamma'}K_1(m-m',\gamma-\gamma')P_{m-m'}^{-1}\Delta W_{\gamma,\gamma'}\frac{m'}{r}{\psi}_{-m',-\gamma'}\Big\rangle\Bigg\},
\end{equation}
\begin{eqnarray}
D_{n\psi}^{(1)}(m,\gamma)&=&\frac{1}{2\pi i}\int_{-i\infty+\gamma_0}^{i\infty+\gamma_0}d\gamma'\sum_{m'\neq0,m}\Bigg\{\frac{m'}{r}\Big\langle\tilde\psi_{m',\gamma'}K_1(m-m',\gamma-\gamma')\nabla^2P_{m-m'}^{-1}\Delta W_{\gamma,\gamma'}\frac{m'}{r} \times \nonumber \\ [2mm]
\tilde{\psi}_{-m',-\gamma'}\Big\rangle&-&\frac{(m-m')}{r}\Big\langle\frac{\partial \tilde{\psi}_{m',\gamma'}}{\partial r}K_1(m-m',\gamma-\gamma')P_{m-m'}^{-1}\Delta W_{\gamma,\gamma'}\Big(\frac{m}{r}\Big)\frac{\partial \tilde{\psi}_{-m',-\gamma'}}{\partial r}\Big\rangle \nonumber \\ [2mm]
&+&\frac{m}{r}\Big\langle K_1(m-m',\gamma-\gamma')P_{m-m'}^{-1}\Delta W_{\gamma,\gamma'}\frac{m'}{r}\tilde{\psi}_{-m',-\gamma'}\nabla^2\tilde{\psi}_{m',\gamma'}\Big\rangle \nonumber \\ [2mm]
&+&\frac{m'}{r}\Big\langle \tilde{\psi}_{m',\gamma'}K_1(m-m',\gamma-\gamma')P_{m-m'}^{-1}\Delta W_{\gamma,\gamma'}\frac{m'}{r}\nabla^2\tilde{\psi}_{-m',-\gamma'}\Big\rangle \nonumber \\ [2mm]
&+&\frac{m'}{r}\Big\langle \tilde{\psi}_{m',\gamma'}K_3(m-m',\gamma-\gamma')\Delta W_{\gamma,\gamma'}\frac{m'}{r}\nabla^2\tilde{n}_{-m',-\gamma'}\Big\rangle \Bigg\},
\end{eqnarray}
where
\begin{equation}
K_2(m,\gamma)=-\Big[1-P_m^{-1}\Big(\hat{\gamma}-D_{nn}^{(1)}(m,\gamma)\frac{\partial^2}{\partial r^2}\nabla_{m}^2-D_{nn}^{(2)}(m,\gamma)\frac{\partial^2}{\partial r^2}\Big)\Big]\Big(imj'(r-r_0)+D_{\psi n}(m,\gamma)\frac{\partial^2}{\partial r^2}\Big)^{-1},
\end{equation}
\begin{equation}
K_3(m,\gamma)=K_1(m,\gamma)K_2(m,\gamma)+\Big(-\hat{\gamma}+D_{\psi\psi}(m,\gamma)\frac{\partial^2}{\partial r^2}\nabla_{m}^2-\frac{im}{r}\frac{d}{dr}n_0\Big)^{-1},
\end{equation}
and $P_m$ and $K_1(m,\gamma)$ are given in Eqs.~(26) and (27).  These expressions contain both linear wave terms and nonlinear diffusivities, and are valid in both weak and strong turbulence regimes.  Outside filaments, where turbulence levels are evaluated to derive the strong refraction condition, Eq.~(37), the turbulence is strong.  The strong turbulence limit of the above expressions yields the diffusivity $d$ that appears in Eq.~(37).

\begin{figure}
\plotone{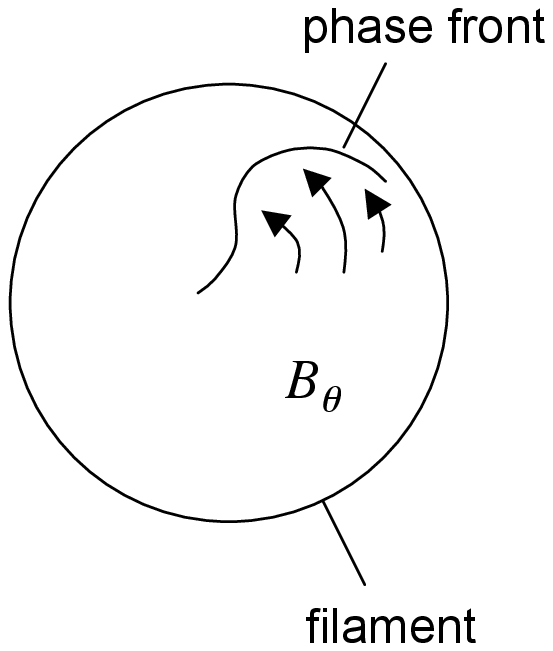}
\caption{Distortion of a kinetic Alfv\'{e}n wave phase front by a sheared filament field.}
\end{figure}

\begin{figure}
\plotone{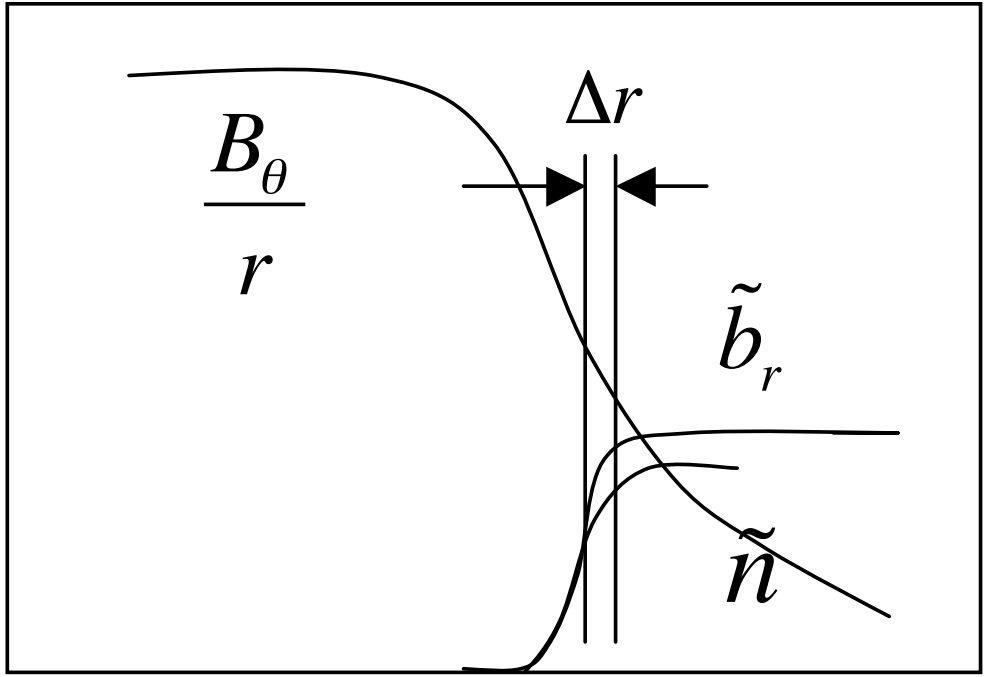}
\caption{Boundary layer at edge of filament (just inside $r=a$).  External fluctuations rapidly decay across the layer.}
\end{figure}

\begin{figure}
\plotone{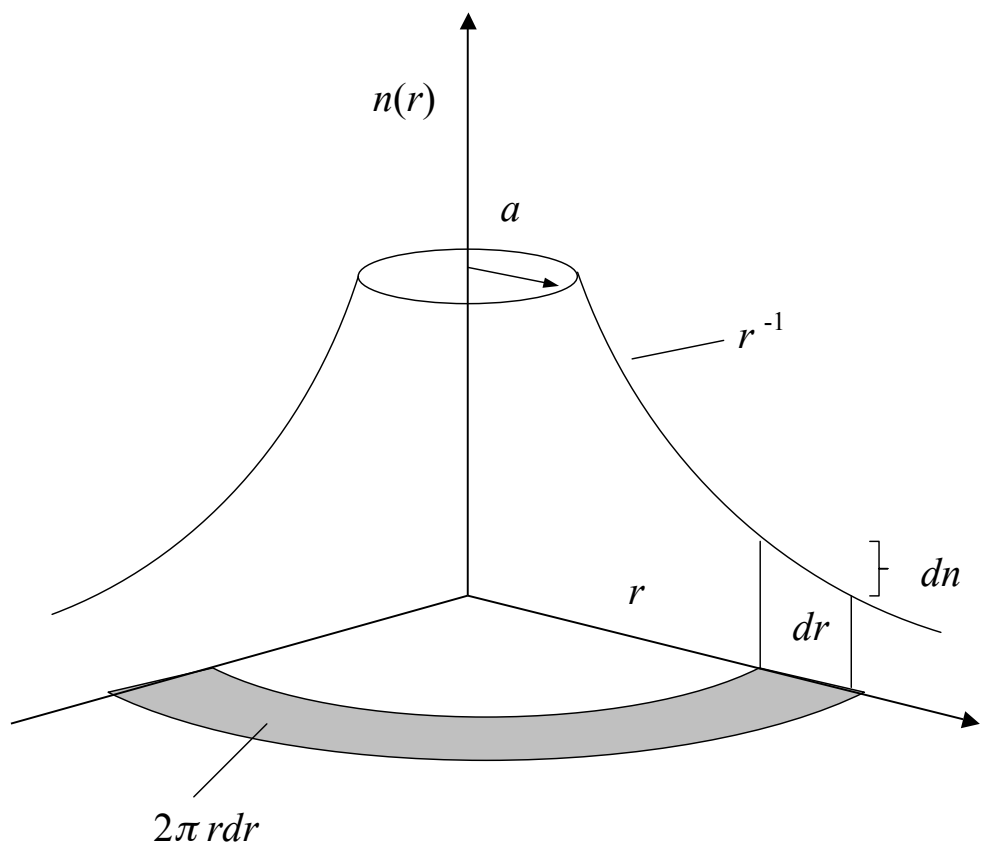}
\caption{The probability of density outside $r=a$ maps onto the annular area $2\pi rdr$.}
\end{figure}

\end{document}